%Paper: hep-ph/9310317
%From: FTLIPKIN@WEIZMANN.weizmann.ac.il
%Date: 20 Oct 93   12:45 +02

%macropackage=phyzzx
\vsize=7.5in
\hsize=6.6in
\hfuzz=20pt%this is because one equation is a bit wide.
\tolerance 10000

\baselineskip 12pt plus 1pt minus 1pt
\pageno=0

\def\endlist{\par}

\headline={\ifnum\pageno=1\firstheadline\else
\ifodd\pageno\rightheadline \else\leftheadline\fi\fi}
\def\firstheadline{\hfil}
\def\rightheadline{\hfil}
\def\leftheadline{\hfil}
        \footline={\ifnum\pageno=1\firstfootline\else\otherfootline\fi}
\def\firstfootline{\rm\hss\folio\hss}
\def\otherfootline{\hfil}
\font\tenbf=cmbx10
\font\tenrm=cmr10
\font\tenit=cmti10
\font\elevenbf=cmbx10 scaled\magstep 1
 1
\font\elevenit=cmti10 scaled\magstep 1

%\TagsOnRight
\nopagenumbers
\line{\hfil }
\vglue 1cm
\hsize=6.0truein
\vsize=8.5truein
\parindent=3pc
\baselineskip=10pt
\centerline{\tenbf A SYMMETRY APPROACH TO CP VIOLATION}
\vglue 5pt
\vglue 1.0cm
\centerline{\tenrm
HARRY J. LIPKIN
}
\baselineskip=13pt
\centerline{\tenit
Department of Physics, Weizmann Institute of Science
}
\baselineskip=12pt
\centerline{\tenit
Rehovot, Israel
}
\vglue 0.8cm
\centerline{\tenrm ABSTRACT}
\vglue 0.3cm
{\rightskip=3pc
 \leftskip=3pc
 \tenrm\baselineskip=12pt%\parindent=1pc
 \noindent
One of the greatest challenges for particle physics in the 1990's is
understanding the broken symmetry of CP violation. It is now almost 30
years since the discovery in 1964 of the $K_{L} \rightarrow 2\pi$ decay.
What has happened since? Why has there been no significant new experimental
input in this long period? The original $K_{L} \rightarrow 2\pi$ decay
experiment is described by two parameters $\epsilon$ and $\epsilon'$. Today
$\epsilon \approx $ its 1964 value while $\epsilon'$ still consistent with
zero, and there is no new evidence for CP violation outside the kaon system.
Why is it so hard to find CP violation? How can B Physics Help? We present
a symmetry approach to these questions.
\vglue 0.8cm }
\line{\elevenbf 1. Introduction \hfil}
\vglue 0.4cm
\line{\elevenit 1.1.
Two approaches to detection of CP Violation
\hfil}
\smallskip

There are two kinds of experimental phenomena which can exhibit CP violation.

(1) Charge asymmetries between the decays of
charge conjugate states $M^{\pm}$ into charge conjugate exclusive final
states $f^{\pm}$; i.e.
$M^+ \rightarrow f^+ \not= M^- \rightarrow  f^-$.
So far no such charge asymmetries have been found.

(2) CP violation in Neutral Meson Mixing. The two mass eigenstates resulting
from mixing both decay into same the CP eigenstate. This is found in
the neutral kaon system - both mass eigenstates, $K_S$ and $K_L$ decay into
two pions. So far this is the only experimental evidence for CP
violation.$^1$

By using a symmetry approach we can understand why it is so difficult
to observe charge asymmetries, and see the essential features of neutral
meson mixing.

As a guide to experimenters,
symmetries show what works, what doesn't work and why.

\bigskip
\endpage
\line{\elevenit 1.2.
What does $K_L \rightarrow 2\pi$ imply? CP before and after
\hfil}
\smallskip

Before 1964 the two kaon flavour eigenstates  $K^o$ and $\bar K^o$
carrying strangeness $\pm 1$ were known to be produced in strangeness
conserving strong interactions; e.g.
$$ K^++n \rightarrow K^o+p, ~ ~ ~ ~ ~ K^-+p \rightarrow \bar K^o + n
\eqno (1.1a) $$
and believed to go into one another under a conserved $CP$ operation.
$$ CP \ket{K^o} =  - \ket{\bar K^o} \eqno (1.1b) $$
The transition matrix elements for the CP-conserving
$\pi^+ \pi^-$ decay satisfy the relation,
$$ \bra{\pi^+ \pi^-}T\ket{K^o} = - \bra{\pi^+ \pi^-}T\ket{\bar K^o}
\eqno (1.2) $$
The mass eigenstates are the $CP$ eigenstates
$$ \ket{K_1} = (1/\sqrt 2)( \ket{K^o} -  \ket{\bar K^o});
{}~ ~ ~ ~  \tau_1 = 9 \times 10^{-11} sec \eqno (1.3a) $$
$$ \ket{K_2} = (1/\sqrt 2)( \ket{K^o} +  \ket{\bar K^o});
{}~ ~ ~ ~
\tau_2 = 5 \times 10^{-8} sec \eqno (1.3b) $$
The two states have very different lifetimes because the dominant decay mode
with the largest phase space is allowed by $CP$ for $K_1$ and forbidden for
$K_2$.
$$ \bra{\pi^+ \pi^-}T\ket{K_2} =  \bra{\pi^+ \pi^-}T\ket{K^o}
+ \bra{\pi^+ \pi^-}T\ket{\bar K^o}  = 0  \eqno (1.4) $$

The discovery that the long-lived kaon also decayed into $\ket{\pi^+ \pi^-}$
immediately showed $CP$ violation, which was described by defining the
following parameters:
$$ \eta_{+-} \equiv {{\bra{\pi^+ \pi^-}T\ket{K_L}}\over{
\bra{\pi^+ \pi^-}T\ket{K_S}}} \equiv \epsilon + \epsilon'
\approx 2.27 \times 10{-3}  \eqno (1.5a) $$
$$ \eta_{oo} \equiv {{\bra{\pi^o \pi^o}T\ket{K_L}}\over{
\bra{\pi^o \pi^o}T\ket{K_S}}} \equiv \epsilon - 2\epsilon'
\approx 2.25 \times 10{-3}  \eqno (1.5b) $$
$$ {{\epsilon'}\over{\epsilon}} \approx (2.2 \pm 1.1 ) \times 10{-3}
\eqno (1.6) $$
where the numerical values are qualitatively correct but may not be exactly
up to date. The value of $\epsilon'$ is still consistent both with zero and the
value predicted by the standard model.

\bigskip
\endpage
\line{\elevenit 1.3.
Use of the EPR effect in neutral meson mixing.
\hfil}
\smallskip

A linear combination of the two mass eigenstates can always be constructed
for which a given decay mode is forbidden; e.g.
$$ \ket{K_\nu^\pm} = \ket{K_L} - \eta_{+-} \ket{K_S} ; ~ ~ ~
 \bra{\pi^+ \pi^-}T\ket{K_\nu^\pm} = 0  \eqno (1.7a) $$
$$ \ket{K_\nu^{oo}} = \ket{K_L} - \eta_{oo} \ket{K_S} ; ~ ~ ~
 \bra{\pi^o \pi^o}T\ket{K_\nu^{oo}} = 0  \eqno (1.7b) $$
The difference between these two states is
proportional to the parameter $\epsilon'$
$$ \ket{K_\nu^{oo}} - \ket{K_\nu^\pm} = 3 \epsilon' \ket{K_S}  \eqno (1.8) $$
A $\ket{K_\nu^{\pm}}$ beam  should not decay to $\pi^+ \pi^-$, while
the decay $K_\nu^{\pm} \rightarrow \pi^o \pi^o$ is proportional to
$\epsilon'$ and could be used in a null experiment to determine $\epsilon'$.
$$ \bra{\pi^o \pi^o}T\ket{K_\nu^{\pm}} =
- 3 \epsilon' \bra{\pi^o \pi^o}T\ket{K_S}  \eqno (1.9) $$

The Einstein-Podolsky-Rosen effect provides a means for creating a
$\ket{K_\nu^{\pm}}$ beam. Consider the decay of the $\phi$ vector meson
at rest into two neutral kaons with momenta $\vec k$ and $-\vec k$
$$\phi \rightarrow K^o(\vec k) \bar K^o(-\vec k) - K^o(-\vec k)
\bar K^o(\vec k)  \eqno (1.10a) $$
This same wave function can also be written in a basis
$(K_\mu^\pm; K_\nu^\pm)$ where the state $K_\mu^\pm$ is defined to be
orthogonal to the state $K_\nu^\pm$,
$$\phi \rightarrow K_\nu^\pm(\vec k) K_\mu^\pm(-\vec k) - K_\mu^\pm(-\vec k)
\bar K_\nu^\pm(\vec k)  \eqno (1.10b) $$
If a decay $K_\mu^\pm \rightarrow \pi^+ \pi^-$  is detected at $-\vec k$, the
wave function collapses to make $K_\nu^\pm$ beam at $\vec k$.
This proposal was called ``An experiment for the future" when first suggested
in 1968.$^2$
Now we hear suggestions for carrying out such experiments at
$\phi$ factories, and the EPR effect is in common use in $B$ decay experiments
using the $B$ analog of the $\phi$, the first bottomonium state $\Upsilon (4S)$
above $B \bar B$ threshold.

\medskip

\endpage
\line{\elevenbf 2. Detecting Charge Asymmetries in Decays
\hfil}
\vglue 0.4cm
\line{\elevenit 2.1.
How CPT complicates detection of CP Violation
\hfil}
\smallskip

Can decays of $K^+$ and $K^-$ be different? For decays to a pair of charge
conjugate final states $\ket{f^{\pm}}$ described by the Fermi
Golden Rule,
$$ W_{K^{\pm}\rightarrow f}\approx (2\pi/\hbar)
|\bra{f^{\pm}}H_{wk}\ket{K^{\pm}}|^2   \rho(E_f)  \eqno (2.1) $$
But from CPT and hermiticity, we see that there can be no asymmetry,
$$ |\bra{f^-}H_{wk}\ket{K^-}|^2 = |\bra{K^+}H_{wk}\ket{f^+}|^2
= |\bra{f^+}H_{wk}\ket{K^+}|^2 \eqno (2.2a)
 $$
$$ W_{K^+\rightarrow f^+}\approx  W_{K^-\rightarrow f^-} \eqno (2.2b)
 $$
CPT also requires equal total widths of $K^+$ and $K^-$. This is
easily seen by noting that s-wave elastic $\pi^{\pm}\pi^o$ scatterings go into
one another under $CPT$. Thus
$\sigma_{el,s}(\pi^+\pi^o) = \sigma_{el,s}(\pi^-\pi^o)$ in the
neighborhood of the kaon mass and is a very narrow Breit-Wigner resonance with
the same width for both charge states,
$$ \Gamma_{tot}(K^+) = \Gamma_{tot}(K^-)         \eqno (2.3)            $$

\noindent Thus the following conditions are necessary for observation of
charge-asymmetric decays:
\pointbegin Golden rule breaks down. This is exact first order
perturbation theory and can only break down where higher order contributions
are
important. Second-order weak contributions are negligible; thus higher order
strong contributions are needed.
\point Conspiracy of several decay modes. Total widths must be equal.
Any asymmetry in the partial widths of a pair of conjugate modes must be
compensated by opposite asymmetries in other modes.
\endlist

We see immediately that it is difficult to satisfy these conditions in the kaon
system. At the kaon mass s-wave $\pi^{\pm}\pi^o$ scattering can only be
elastic; no inelastic channels are open. Thus the s-wave $\pi^{\pm}\pi^o$ state
is an exact eigenstate of the strong interaction S-matrix, the golden rule
holds
for $K^{\pm} \rightarrow \pi^{\pm}\pi^o$ and no charge asymmetry can be
observed. Some possibilities exist in other decay modes, like $3 \pi$, where
the
$\pi^{\pm} \pi^{\pm} \pi^{\mp}$ and $ \pi^{\pm} \pi^o  \pi^o $ modes are
coupled. However, these are linear combinations of two isospin eigenstates with
I=1 and I=3. The I=3 amplitude is expected to be suppressed; it is a
$\Delta I = 5/2$ transition and doubly suppressed by the
$\Delta I = 1/2$ rule. Thus the
I=1 amplitude is nearly an eigenstate of the strong interaction
S matrix and the golden rule should be a good approximation. A similar
situation obtains for different partial wave amplitudes which are coupled. Here
the overall s-wave is expected to be dominmant and again be an approximate
strong S-matrix eigenstate. Thus all charge asymmetry effects in the kaon
system are expected to be small.

\bigskip
\line{\elevenit 2.2. Beating CPT for Charge Asymmetries in B Physics \hfil}
\smallskip

Can decays of $B^+$ and $B^-$ be different? Here many more channels are open,
different decay modes can conspire to give the same total width and
Final state rescattering can beat the Fermi golden rule
via higher order transitions in strong interactions; e.g.
$$ B^- \rightarrow \bar K^o \pi^- \rightarrow K^-\pi^o ; ~ ~ ~ ~ ~
B^+ \rightarrow K^o \pi^+ \rightarrow K^+\pi^o \eqno (2.4) $$
This has no simple counterpart in the kaon system where the only open
hadronic channels are $2 \pi$ and $3 \pi$ and the I=3 amplitude is strongly
suppressed. Here both $(K\pi)$ isospin eigenstates  $I=1/2$ and $I=3/2$
are produced by $\Delta I = 1$ transitions and are equally allowed.

How a CP-violating asymmetry can be obtained
is very simply illustrated in a toy model where only $K\pi$ decay
modes contribute to $B$ decay. The isospin eigenstates $(K\pi)_I$ where $I=1/2$
and $I=3/2$ are eigenstates of the strong interaction S-matrix and are both
expected to be produced without any suppression. Thus the golden
rule applies to decays into these states. Since the strong interactions are
exactly diagonalized and the higher order weak interactions are negligible
there
are no higher order corrections to decays into isospin eigenstates in this
model. Thus from CPT and hermiticity there can be no charge asymmetry in decays
to isospin eigenstates. For I=1/2 and 3/2,
$$ |\bra{(\bar K\pi)_I}H_{wk}\ket{B^-}|^2 =
|\bra{(K\pi)_I}H_{wk}\ket{B^+}|^2  \eqno (2.5a) $$
$$ \Gamma\{B^+\rightarrow (K\pi)_I\} = \Gamma\{B^-\rightarrow
(\bar K\pi)_I\} \eqno (2.5b)
 $$
Then
$$ \Gamma_{tot}(B^+) = \sum_I \Gamma\{B^+\rightarrow (K\pi)_I\}
= \Gamma_{tot}(B^-) = \sum_I \Gamma\{B^-\rightarrow (\bar K\pi)_I\}
\eqno (2.6)  $$
in agreement with the $CPT$ requirement of charge symmetric total widths.

However, asymmetries can occur for decays into final states which are not
strong interaction eigenstates; e.g. $K^{\pm}\pi^o$:
 $$A\{B^+ \rightarrow K^+\pi^o \} =
 \sum_I C^{+o}_I|A\{B^+ \rightarrow (K  \pi)_I\}|
\cdot e^{iW_I}e^{iS_I} \eqno (2.7a)
 $$
 $$A\{B^- \rightarrow K^-\pi^o \} =
 \sum_I C^{+o}_I |A\{B^+ \rightarrow (\bar K  \pi)_I\}|
\cdot e^{-iW_I}e^{iS_I} \eqno (2.7b) $$
where $ C^{+o}_I $ denotes Clebsch-Gordan coefficients for isospin couplings.
We have written every isospin amplitude as the product of its magnitude, a
weak phase factor $e^{-iW_I}$ and a strong phase factor $e^{iS_I}$, and noted
that the weak CP-violating phase reverses sign under charge conjugation while
the strong CP-conserving phase remains unchanged.  Then
I=3/2 - 1/2 interference can produce charge asymmetry,
$$|A\{B^+ \rightarrow K^+\pi^o \}|^2 - |A\{B^- \rightarrow K^-\pi^o \}|^2 = $$
$$ = C^{+o}_1 C^{+o}_3|A_1A_3|\cdot
\{e^{i(W_1-W_3)}e^{i(S_1-S_3)}-e^{i(W_3-W_1)}e^{i(S_1-S_3)}\}
+c.c.  \eqno (2.8) $$
The asymmetry is seen to vanish unless $both $ $W_1 \not= W_3$ $and$
$S_1 \not= S_3$.
Thus the condition for observing an asymmetry is that at least two amplitudes
must contribute which arise from different strong interaction eigenstates, and
that these amplitudes must have both different strong phases and different weak
phases.
\bigskip
\line{\elevenit 2.3.
Charge Asymmetry in Standard Model - Trees and Penguins
 \hfil}
\smallskip
In the standard model two different diagrams with different weak phases can
contribute to $B \rightarrow K\pi$ decays via two different strong interaction
eigenstates. There is therefore a possibility of observing a CP asymmetry.

\noindent The tree diagram gives only the $K^{\pm}\pi^o$ final states.
$$B^+(\bar b u) \rightarrow \bar u + W^+ + u \rightarrow \bar u + u +
\bar s + u \rightarrow K^+ + \pi^o  \eqno (2.9a) $$
$$B^-(b \bar u) \rightarrow u + W^- +\bar u \rightarrow u + \bar u + s +
\bar u \rightarrow K^- + \pi^o  \eqno (2.9b) $$
The penguin diagram gives only the I=1/2 $K-\pi$ final state.
$$B^+(\bar b u) \rightarrow \bar t + W^+ + u \rightarrow  \bar s + u
\rightarrow (K  \pi)_{I=1/2} \eqno (2.10a) $$
$$B^-(b \bar u) \rightarrow t + W^- +\bar u \rightarrow s + \bar u
\rightarrow  (\bar K  \pi)_{I=1/2} \eqno (2.10b) $$
In this case the tree contribution is strongly suppressed. It involves both
suppressed weak vertices, $b \rightarrow u$ and $s \rightarrow u$. There is
therefore hope that tree-penguin interference may be
observed. So far no penguin contributions have been unambiguously identified.

\medskip
\endpage
\vglue 0.6cm
\line{\elevenbf 3. Symmetry Analysis of Neutral Meson $M^o - \bar M^o$ Mixing
\hfil}

\vglue 0.4cm

\line{\elevenit 3.1.
A Quasispin Description of Neutral Meson Mixing
\hfil
}
\smallskip

It is convenient to describe neutral meson mixing by a quasispin SU(2) picture
in which the two flavor eigenstates of the meson system, denoted by $M$ and
$\bar M$ are defined as eigenstates of $\sigma_z$ with ``spin up" and "spin
down" respectively. Thus
$$ \sigma_z \ket{M^o} = \ket{M^o} ; ~ ~ ~  \sigma_z \ket{\bar M^o} =
- \ket{\bar M^o}  \eqno (3.1)  $$
Strong and electromagnetic interactions conserve quasispin.
Weak interactions break quasispin.

If CPT is conserved the mass eigenstates $M_S$ and $M_L$ are equal mixtures of
$M^o$ and $\bar M^o$.  We can then choose quasispin x axis to make
$$ \sigma_x \ket{M_S} = \ket{M_S} ; ~ ~ ~ \sigma_x \ket{M_L} = - \ket{M_L}
\eqno (3.2) $$
If CP is conserved, $M_S$ and $M_L$ are also CP eigenstates.

If CP is violated, $M_S$ and $M_L$ can both decay into the same given
CP eigenstate $\ket f$. But a basis $(M_\nu;M_\mu)$ can be defined to make
   $ \bra{f} H_{weak} \ket {M_\nu}  =     0 $.
If $M_\nu$ and $M_\mu$ also equal mixtures of $M^o$ and $\bar M^o$, as occurs
in many cases, they define a direction in the $x-y$ plane at some angle
$\theta_f$ with the
$x$ axis. The values of $\theta_f$ for two different CP eigenstates are
directly to the CP-violation parameters $\epsilon$ and $\epsilon'$.
If $CP$ is conserved, $\theta_f = 0$.
\vglue 0.6cm
\bigskip
\line{\elevenit 3.2.
Quasispin Symmetry Breaking by Weak Interaction
\hfil
}
\smallskip
There are two symmetry-breaking mechanisms.

Breaking by the lifetime difference is
dominant in the kaon system where a pure $K_L$ state can be produced simply by
waiting. The breaking is determined by phase space and independent of the
Standard model. Lifetime breaking is negligible in heavy quark mesons.

Breaking by the mass difference is dominant in heavy quark mesons
and determined by dynamical effects depending upon standard model.
In the quasispin picture this breaking can be described as a ``magnetic field"
in the quasispin space. The time dependence of the heavy meson mixing is then
described as a quasispin precession in the magnetic field.

Experiments can be described as the production of a quasispin polarized beam
followed by a subsequent polarization measurement.

The time development of a general neutral $B$ meson which is in the state
$\ket{B(0)}$ at time $t=0$ is given by
$$\ket{B(t)} = e^{-{\Gamma\over 2} t} e^{-i\omega \sigma_x (t/2)}\ket{B(0)}
= e^{-{\Gamma\over 2} t} \cdot\{\cos
({{\omega t}\over 2}) -i \sigma_x \sin ({{\omega t}\over 2})\}
\ket{B(0)} \eqno (3.3) $$
where $\Gamma$ is the decay width and $\omega$ the mass difference between
the two eigenstates.
Then for states which are initially $\ket{B^o}$ and $\ket{\bar B^o}$ at t=0,
$$
\ket{B^o(t)} = e^{-{\Gamma\over 2} t} \cdot\{\cos
({{\omega t}\over 2})\ket{B^o}
-i \sin ({{\omega t}\over 2}) \ket{\bar B^o} \}  \eqno (3.4a) $$
$$  \ket{\bar B^o(t)} = e^{-{\Gamma\over 2} t}
 \cdot\{\cos ({{\omega t}\over 2})\ket{\bar B^o}
- i \sin ({{\omega t}\over 2})\ket{B^o}\} \eqno (3.4b) $$
 Then
$$ 2e^{{\Gamma} t} |\langle B^o \ket{B^o(t)}|^2 =
\cos^2 ({{\omega t}\over 2}) =(1/2)\{1 + \cos (\omega t)\}
\eqno (3.5a) $$
$$ 2e^{{\Gamma} t} |\langle B^o  \ket{\bar B^o(t)}|^2 =
\sin^2 ({{\omega t}\over 2}) =(1/2)\{1 - \cos (\omega t)\}
\eqno (3.5b) $$
This is just the well known $B^o - \bar B^o$ mixing independent of all $CP$
Violation.

\bigskip
\line{\elevenit 3.3.
Experiments as Quasispin Polarization Measurements
\hfil
}
\smallskip
Consider an experiment in which a $B$ meson is prepared in some state
$\ket{B_\nu}$ and its decay is observed after a time $t$
in a mode allowed only for $B$ or allowed only for $\bar B$; e.g. leptonic
modes. The difference between the probabilities of decay into $B$ or $\bar B$
allowed modes; e.g. a lepton asymmetry, is just given by
the quasispin polarization in the $z$ direction of the prepared state.
This is easily evaluated using the Pauli spin algebra.
$$  |\bra{B^o} e^{-i\omega \sigma_x (t/2)} \ket{B_\nu}|^2 -
|\bra{\bar B^o} e^{-i\omega \sigma_x (t/2)} \ket{B_\nu}|^2 =
\bra{B_\nu}  e^{i\omega \sigma_x (t/2)} \sigma_z
 e^{-i\omega \sigma_x (t/2)} \ket{B_\nu} $$
$$ = \bra{B_\nu}  \sigma_z
 e^{-i\omega \sigma_x t} \ket{B_\nu}
 = \bra{B_\nu}  \sigma_z \cos (\omega t) - i \sigma_z \sigma_x
\sin (\omega t) \ket{B_\nu} = $$
$$ = \bra{B_\nu}  \sigma_z \ket{B_\nu} \cos (\omega t) +
\bra{B_\nu}  \sigma_y  \ket{B_\nu} \sin (\omega t) =
\sin (\omega t) \sin \theta \eqno (3.6) $$
where the last equality holds for the case where the state $\ket{B_\nu}$ is an
equal mixture of $B^o$ and $\bar B^o$ and thus
has its quasispin in the direction of an axis in the $x-y$ plane. The angle
$\theta$ between this axis and the $x$ axis is determined by the relative phase
of the $B^o$ and $\bar B^o$ components. This situation occurs often in
experiments where the state $\ket{B_\nu}$ is prepared by observing a decay into
a $CP$ eigenstate. The angle $\theta$ then gives a measure of the $CP$
violation.

A common example of such an experiment is one in which a neutral
$B \bar B$ pair is created by the decay of the $\Upsilon (4S)$, one $B$ decays
in the $K_S \psi$ decay mode, and the other decays into a leptonic mode.
Let us define a
basis $(B_\mu;B_\nu)$ to make $ \bra{K_S \psi}T\ket{B_\nu} = 0$.
The second $B$ is required to be in the state $ \ket{B_\nu}$ at the time that
the $K_S \psi$ decay of the other $B$ is observed. The lepton asymmetry
observed
in the second decay at a time $t$ after the first decay is seen to be given by
the expression (3.6). Another interesting identity relevant to this experiment
is obtained by use of the quasispin algebra:
$$
|\bra{B_\mu}e^{-i\omega \sigma_x (t/2)}\ket{B^o}| =
|\bra{B_\mu}e^{-i\omega \sigma_x (t/2)}\sigma_z \ket{B^o}|
 = |\bra{B_\mu}\sigma_z e^{i\omega \sigma_x (t/2)} \ket{B^o}| =  $$
$$ =
|\bra{B_\nu} e^{i\omega \sigma_x (t/2)} \ket{B^o}| =
|\bra{B^o} e^{-i\omega \sigma_x (t/2)} \ket{B_\nu}^*|
\eqno (3.7) $$
Thus the probability that a meson created as a $B^o$ at time $t_1$
will be observed as a $B_\mu$ at time $t_2$ is exactly equal to
the probability that a meson created as a $B_\nu$ at time $t_1$
will be observed as a $B^o$ at time $t_2$.
$$ P\{B^o(t_1) \rightarrow B_\mu(t_2)\} =
P\{B_\nu(t_1) \rightarrow B^o(t_2)\}
\eqno (3.8) $$
We now see the implications of this identity for the $\Upsilon(4S)$ experiment
in which one $B \rightarrow K_S \psi$ and the other decays leptonically.
$$
P\{\Upsilon(4S) \rightarrow \bar B^o(t_1) B_\mu(t_2)\} =
P\{\Upsilon(4S) \rightarrow \bar B^o(t_1) B^o(t_1)\}\cdot
P\{B^o(t_1) \rightarrow B_\mu(t_2)\}
\eqno (3.9a) $$
$$
P\{\Upsilon(4S) \rightarrow B_\mu(t_1) B^o(t_2)\} =
P\{\Upsilon(4S) \rightarrow B_\mu(t_1) B_\nu(t_1)\} \cdot
P\{B_\nu(t_1) \rightarrow B^o(t_2)\}
\eqno (3.9b) $$
$$ P\{\Upsilon(4S) \rightarrow \bar B^o(t_1) B_\mu(t_2)\} =
P\{\Upsilon(4S) \rightarrow B_\mu(t_1) B^o(t_2)\}
\eqno (3.10) $$
The lepton asymmetry observed at time $t_1$ when a $K_S \psi$ decay is
observed at time $t_2$ is seen to be exactly equal and opposite to
lepton asymmetry observed at time $t_2$ when a $K_S \psi$ decay is
observed at time $t_1$. Thus in this kind of experiment the CP-violating lepton
asymmetry cancels out if the results are integrated over time. Since time
measurements are difficult in the rest frame of the $\Upsilon(4S)$ where the
two $B$ mesons move very slowly ``asymmetric B factories" have been proposed
to produce the $\Upsilon(4S)$ in flight so that the $B$ mesons traverese a
measurable distance before decay.

\medskip

\endpage
\vglue 0.6cm
\line{\elevenbf 4. Summary - How $B$ and $K$ physics differ -
Good and bad news
\hfil}

\vglue 0.4cm

\pointbegin No Dominant $B$ Decay Mode
\spoint No Lifetime Difference
\spoint Mass Eigenstates Not Separated by Waiting
\endlist
\point Many $B$ Decay Modes
\spointbegin Rich Data - Small Branching Ratios $\approx$ 1\%
\spoint Final State Rescattering - Beats Golden Rule
\spoint Conspiracies Beat CPT Restrictions
\endlist
\point $B^o - \bar B^o$ Oscillations During Decay
\spointbegin Time Dependence Confuses Measurements
\spoint CP Violation Observable in Mixing Phases
\endlist
\point All Dominant Hadronic B decays involve 3 Generations
\spointbegin CP violation Observable in B Decays in Direct Diagrams
$ b \rightarrow c \bar u d     $
\spoint CP Violation Observable in charm and strangeness decays only via
diagrams with virtual t and b quarks
\endlist
\endlist
\vglue 0.6cm
\line{\elevenbf 5. References \hfil}
\vglue 0.4cm
\item{1.} J.H. Christensen et al., {\elevenit Phys. Rev. Lett.}
{\elevenbf 13} (1964) 138
\item{2.} Harry J. Lipkin, {\elevenit Phys. Rev. }
{\elevenbf 176} (1968) 1715
\vglue 0.6cm

\eject
\bye
\end